# Monolayer Phosphorene — Metal Interfaces

Yuanyuan Pan[1], Yangyang Wang[1,3], Meng Ye[1], Ruge Quhe[1,4], Hongxia Zhong[1], Zhigang Song[1], Xiyou Peng,[1] Dapeng Yu[1,2], Jinbo Yang[1,2], Junjie Shi[1], Ju Li,[3] and Jing Lu[1,2,*]

[1]State Key Laboratory of Mesoscopic Physics and Department of Physics, Peking University, Beijing 100871, P. R. China

[2]Collaborative Innovation Center of Quantum Matter, Beijing 100871, P. R. China

[3]Department of Nuclear Science and Engineering and Department of Materials Science and Engineering, Massachusetts Institute of Technology, Cambridge, Massachusetts 02139 , USA

[4]Academy for Advanced Interdisciplinary Studies, Peking University, Beijing 100871, P. R. China

[*]Corresponding author: jinglu@pku.edu.cn

**Abstract**

Recently, phosphorene electronic and optoelectronic prototype devices have been fabricated with various metal electrodes. We systematically explore for the first time the contact properties of monolayer (ML) phosphorene with a series of commonly used metals (Al, Ag. Cu, Au, Cr, Ni, Ti, and Pd) via both *ab initio* electronic structure calculations and more reliable quantum transport simulations. Strong interactions are found between all the checked metals, with the energy band structure of ML phosphorene destroyed. In terms of the quantum transport simulations, ML phosphorene forms a *n*-type Schottky contact with Au, Cu, Cr, Al, and Ag electrodes, with electron Schottky barrier heights (SBHs) of 0.30, 0.34, 0.37, 0.51, and 0.52 eV, respectively, and *p*-type Schottky contact with Ti, Ni, and Pd electrodes, with hole SBHs of 0.30, 0.26, and 0.16 eV, respectively. These results are in good agreement with available experimental data. Our findings not only provide an insight into the ML phosphorene-metal interfaces but also help in ML phosphorene based device design.

Two-dimensional (2D) materials with atomic thickness, such as graphene[1] and transition metals dichalcogenides[2] (TMDCs), have attracted numerous attentions, due to their excellent properties and great potential for electronic and optoelectronic applications,[3,4,5,6,7,8]. However, the zero band gap of graphene limits its application in digital logic devices despite of an extremely high carrier mobility of about 20000 cm V$^{-1}$ s$^{-1}$, and the relatively low mobility (200 cm V$^{-1}$ s$^{-1}$) of TMDCs (with a band gap of 1-2 eV) is difficult to meet the future request of high frequency applications. Phosphorene, the youngest member of 2D materials,[9,10] has a finite band gap from 0.3 (in bulk) to 1.0 (monolayer) eV[11] and is mechanically exfoliated in the early of 2014[12,13,14,15]. Monolayer (ML) and few-layer (FL) phosphorene field effect transistors (FET) have been successfully fabricated with on-off ratio of 10$^5$ and a mobility up to 1000 cm V$^{-1}$ s$^{-1}$, which make phosphorene a competitive candidate channel material for future electronic applications. FL phosphorene photo-detector has been realized by Engel *et al.*, which is capable of acquiring high-contrast (visibility V > 0.9) images both in the visible as well as in the infrared spectral regime.[16] FL phosphorene phototransistor was also firstly demonstrated with a rise time of about 1 ms and responsivity of 4.8 mA/W.[17] Phosphorene *p-n* diode has been proposed based on *p*-type phosphorene and *n*-type MoS$_2$ or graphene van der Waals heterojunction.[18,19,20] Therefore, phosphorene is anticipated a promising material in photoelectronic devices.[21,22]

In an actual device, a contact with metal is usually required in order to inject appropriate types carrier into the conduction or valence band of semiconducting 2D materials, and the formation of low-resistance metal contacts is the biggest challenge that masks the intrinsic exceptional electronic properties of semiconducting 2D materials[23]. In the absence of controllable and sustainable substitutional doping scheme in 2D materials, a contact with metal also controls the carrier type of semiconducting 2D materials. In such metal semiconductor contact, Schottky barrier is often formed, which decreases electron or hole injection efficiency. In order to obtain a high performance of a device, one can select a metal electrode with a small Schottky barrier height (SBH) at the metal-semiconductor interface. Unfortunately, the SBH is not simply determined by the difference between the work function of a metal and the conduction band minimum (CBM) or the valence band maximum (VBM) of semiconducting 2D materials due the complex Fermi level pinning.[24]

In experiment, the SBHs of ML phosphorene FET with Ni electrode is measured to be 0.63

± 0.02 eV for holes and 0.35 ± 0.02 eV for electrons.[14] On the theoretical aspect, ML phosphorene contacts with Ta, Nb, Cu, Zn, In, (by Gong et al[25]) Au, Ti, and Pd (by Chanana et al[26]) have been examined by using density functional theory (DFT) band calculations and further by using quantum transport simulation (only for Cu and Zn contacts). They predicted that Cu forms an *n*-type and Pd forms a *p*-type Ohmic contact with ML phosphorene, respectively. The rest metals form Schotty contact with ML phosphorene. These calculations suffer from four shortcomings: (1) the Schottky barrier heights (SBHs) are not calculated. (2) They only considered the Schottky barrier in the vertical interfaces, ignoring that in the lateral interface, and it is proved that such a single interface model fails to reproduce the observed SBHs in ML $MoS_2$ transistors and a dual interface model is required. In the dual interface model of a 2D material transistor (Fig. 6(a)), one interface is the source/drain interface (B) between the contacted 2D material and the metal surface in the vertical direction, and the other is source/drain-channel (D) interface between the contacted 2D material and channel material in the lateral direction as shown in Fig. 6(b).[23,27] (3) The interfaces between the commonly used metals Ni, Cr, Ag, and Al and ML phosphorene has not been examined. (4) Most of the contact properties except for Cu and Zn are derived from the energy band analysis, which, however, ignore the interactions of metal electrodes and channel ML phosphorene and probably leads to artificial Ohmic contact.[27] To correct such a critical deficiency, *ab initio* quantum transport simulation is required.

In this Letter, we systematically explore the interfacial properties between ML phosphorene and the common metals (Al, Ag, Au, Cu, Ti, Cr, Ni, and Pd) based on *ab initio* electronic structure calculations and quantum transport simulations for the first time. According to the bonding level and the hybridization degree of ML phosphorene energy band structure, two classes are revealed: weak chemical bonding is formed for ML phosphorene on Al, Ag, Cu, and Au electrodes with smaller binding energy and farther interlayer distance, while strong chemical bonding is formed for ML phosphorene on Ti, Ni, Cr, and Pd with larger binding energy, closer interlayer distance, and destroy of ML phosphorene configuration. Quantum transport simulations often give quite different SBHs compared with the energy band analysis and are better consistent with the experiments. For example, the calculated (from the transport simulation) and observed *n* type SBH between ML phosphorene and Ni electrodes are 0.26 and 0.35 ± 0.02 eV, respectively. No Ohmic contacts are found between ML phosphorene and the eight metal interfaces.

We consider eight metals Al, Ag, Au, Cu, Ni, Ti, Pd, and Cr, which cover a wide range of workfunction. We choose six layers of metal atoms to simulate the metal surfaces and construct a supercell with ML phosphorene absorbed on one side of the metal surfaces, as shown in Fig. 1c. We fix the lattice constants of metal surface and change the optimized lattice constant of ML phosphorene $a = 3.30$ Å and $b = 4.63$ Å (the lattice constant are indicated in Figure 1) to adjust to them. The $\sqrt{a^2+b^2} \times \sqrt{(2a)^2+b^2}$ ML phosphorene matches $2 \times 2$ Al, Ag, Au, and Cr surface in the (110) orientation, and the $3a \times b$ ML phosphorene match $1 \times 2$ Ti surface in the (110) orientation, respectively, and the $3a \times b$, the $3a \times b$, and the $5a \times b$ ML phosphorene match $2 \times \frac{\sqrt{3}}{2}$ Ni surface, $4 \times \sqrt{3}$ Cu surface and $6 \times \sqrt{3}$ Pd surface in the (111) orientation, respectively. The corresponding mismatches of lattice constant are 0.59 ~ 2.65%, as given in Table 1. A vacuum buffer space of at least 12 Å is set. ML phosphorene mainly interacts with the top layer metal atoms, so the bottom three layers of metal atoms and the cell shape are fixed.

The geometry optimizations and electronic structure calculations are performed with the projector-augmented wave (PAW) pseudopotential[28] and plane-wave basis set with a cut-off energy of 400 eV, implemented in the Vienna *ab initio* simulation package (VASP) code.[29,30] The maximum residual force during geometry optimization is less than 0.01 eV/Å and energies are converged to within $1 \times 10^{-5}$ eV per atom. The Monkhorst-Pack *k*-point mesh is sampled with a separation of about 0.02 Å$^{-1}$ in the Brillouin zone during the relaxation and electronic calculation periods.[31] Van der Waals interaction is taken into account, with the vdW-DF level of optB88 exchange functional (optB88-vdW).[32] Since the slab is not symmetric, a dipole correction is used to eliminate the spurious interaction between the dipole moments of periodic images in the *z* direction. The total electron density are calculated by using ultrasoft pseudopotential plane-wave method implemented in CASTEP code, with a plane-wave cut-off energy of 280 eV.[33]

Transmission spectra and local device density of state (LDDOS) are calculated by using DFT coupled with nonequilibrium Green's function (NEGF) method, which are implemented in Atomistix Tool Kit (ATK) 11.2 package.[34,35,36] Single-$\zeta$ plus polarization (SZP) basis set is employed, the real-space mesh cutoff is at least of 400 eV, and the temperature is set at 300 K. The electronic structures of electrodes and central region are calculated with a Monkhorst–Pack[31] 50 × 1 × 50 and 50 × 1 × 1 $k$-point grid, respectively. GGA of PBE form[37] to the exchange-correlation functional is used.

The most stable configurations of ML phosphorene-metal systems after relaxation are depicted in Figure 2. We have optimized two different initial configurations of ML phosphorene-Al systems, and obtained the same stable configuration with the two adjacent phosphorus atoms of a layer almost on the two adjacent Al atoms of A layer from the side view as shown in Figure 1c. The most configurations of ML phosphorene on Ag, Au, and Cu electrodes are similar with ML phosphorene-Al systems, which is in line with previous work.[25] For ML phosphorene on Al, Ag, Au, and Cu electrodes, the configuration of ML phosphorene is preserved, while it is destroyed seriously for ML phosphorene on Cr, Ni, Ti, and Pd electrodes, so the initial configurations of ML phosphorene on Cr, Ni, Ti, and Pd interface have little effect on relaxed configuration.

The primary parameters of ML phosphorene-metal contacts are listed in Table 1. The binding energy $E_b$ of the ML phosphorene-metal contact is defined as

$$E_b = (E_P + E_M - E_{P\text{-}M}) / N \qquad (1)$$

where $E_P$, $E_M$, and $E_{P\text{-}M}$ are the relaxed energy for pristine ML phosphorene, the clean metal surface, and the combined system, respectively, and $N$ is the number of interfacial a layer phosphorus atoms in a supercell as shown in Figure 1. The equilibrium interlayer distance $d_{P\text{-}M}$ is defined as the average distance from the innermost layer of metal to the innermost ML phosphorene surface in the vertical direction of the interface as shown in Figure 1c. The minimum interatomic distance $d_{min}$ is defined as the minimum atomic distance from the innermost layer atom of metal to ML phosphorene surface atom. According to the bonding level, two categories of ML phosphorene-metal interfaces are revealed after relaxation. Weak bonding is formed for ML phosphorene-Al, Ag, Au, and Cu interfaces with smaller binding energy of 0.41 < $E_b$ < 0.59 eV, and larger interlayer distance of 2.35 < $d_{P\text{-}M}$ < 2.57 Å and 2.30 < $d_{min}$ < 2.46 Å, while strong bonding is formed for ML phosphorene-Cr, Ni, Pd, and Ti interfaces with lager binding energy of 0.85 < $E_b$ < 1.62 eV, and smaller distance of 1.60 < $d_{P\text{-}}$ $_M$ < 2.19 Å, and 2.07 < $d_{min}$ < 2.32 Å. Compared with the elemental single-layer plane carbon material, such as graphene and graphdiyne, the nonplanar single-layer phosphorus material, ML phosphorene is more active. Therefore, the interactions between ML and metals are stronger than those between graphene/graphdiyne and metals with ten times larger binding energy and smaller interfacial distance, which are consistent with that fact the adsorption energies of adatoms on ML phosphorene are more than twice larger than on graphene.[24,38,39,40]

As shown in Figure 3, the band structures of ML phosphorene are destroyed on all the metal electrodes, suggestive of a covalent or chemical bonding between ML phosphorene and these metals. The band hybridization of ML phosphorene on Ti, Ni, Cr, and Pd electrodes is more intense than that on Al, Ag, Cu, and Au electrodes, a fact in line with their bonding level. By contrast, the band structures of ML graphene and graphdiyne absorbed on Ag, Al, Au, and Cu substrates not destroyed. This difference once more indicates that the interactions of ML phosphorene on metal surfaces are fiercer than those of graphene and graphdiyne on metal surfaces.[24,38,39]

The total electron distribution of ML phosphorene-metal adsorbed systems can reveal the interaction at the ML phosphorene-metal interface. We calculated the electron density in the real space of ML phosphorene-Ag, Au, Pd, and Ni adsorbed systems, and cut a slice with the most interfacial atoms shown in Figure 4. There is apparent electron accumulation in the ML phosphorene-Ag, Au, Ni, and Pd interfaces, indicating the formation of a covalent bond between them once more. The electron accumulation at ML phosphorene-Ni and Pd interfaces is more apparent than that at ML phosphorene-Ag and Au interfaces, as are in line with their bonding and hybridization levels.

The schematic diagram of a ML phosphorene FET is shown in Figure 6(a). Schottky barriers can appear on either of the two different interfaces in a ML phosphorene FET: One is between ML phosphorene and the contacted metal surface in the vertical direction (labeled interface B, and the corresponding SBH is labeled $\Phi_V$), and the other is between the contacted systems and the channel ML phosphorene in the lateral direction (labeled interface D, and the corresponding SBH is labeled $\Phi_L$).[41] Besides, tunneling barrier can appear at interface B when electrons cross the gap (normally van der Waals gap) between metal and ML phosphorene. The frequently-used method to investigate contact barriers at 2D semiconductor-metals interfaces is electronic energy band calculation, in which contact barriers is decide by the distribution of the Fermi level ($E_f$) of metals and band structure of semiconductors. From Figure 3, the band hybridizations between ML phosphorene and all checked metals lead to a metallization of ML phosphorene under metal, and vertical Schottky barrier is absent.

Lateral electron SBH $\Phi_L^e$ (hole SBH $\Phi_L^h$) are determined by the energy differences between the interfacial system $E_f$ and the CBM (VBM) of channel ML phosphorene. Apparently, the size of the band gap is critical to evaluate the SBH of a 2D semiconductor device. There are four kinds of commonly used band gap for a 2D semiconductor: transport gap (a sum of the electron SBH and hole SBH), quasi-particle band gap (dominated by many-electron effects), optical gap (dominated by exciton effects), and DFT band gap (single-electron approximation). For ML phosphorene, the four kinds of band gap are 1.0,[14] 2.0,[42] 1.3-1.45,[13,43] and 0.91 (PBE functional) eV, respectively. Obviously, the DFT band gap is closest to the transport SBH because many-electron effects have been strongly suppressed due to charge doping of channel ML phosphorene by metal source/drain electrode or gate electrode. Hence, single-electron approximation is a good approximation to characterize the transport gap and SBH. The same case occurs in ML and BL MoS$_2$-Ti contact, too. The experimentally extracted SBHs of ML and BL MoS$_2$-Ti contact are 0.3 ~ 0.35[23] and 0.065 eV[44], respectively, which are in agreement with calculated values of 0.216 and 0.161 eV at the DFT level (PBE functional).[27]

ML phosphorene forms an Schottky contacts with Al, Ag (*n*-type), and Ti (mid-gap), Cr, Au, Cu, and Ni (*p*-type) electrodes in the lateral direction, with lateral electron SBHs $\Phi_L^e$ = 0.23, 0.30, 0.40, 0.58, 0.71, 0.76, and 0.89 eV, respectively, while forms an Ohmic contact with Pd electrodes since the absorbed system $E_F$ is lower than VBM of channel ML phosphorene by electronic energy calculations. In experiments, it is not easy to form perfect Ohmic contact in 2D semiconductor FET, so there is doubt of the Ohmic contact with Pd electrodes in ML phosphorene.

The potential profiles at the vertical ML phosphorene-metal interfaces are shown in Figure 2. There are obvious tunneling barrier at the ML phosphorene-Cu, Al, Au, and Ag interfaces with the barrier height of 0.65, 1.26, 1.55, and 1.67 eV, respectively, while there are vanishing tunneling barrier at the ML phosphorene-Cr, Ni, Ti, and Pd interfaces since the metallization of them. We assume a square potential barrier to replace the real potential barrier, and the barrier height ($\Delta V$) and width ($w_B$) of the square potential barrier are the barrier height and full width at half maximum (FWHM) of the real potential barrier shown in Fig. 2. The tunneling probabilities $T_B$ is calculated using equation:[45]

$$T = \exp\left(-2\frac{\sqrt{2m\Delta V}}{\hbar} \times w_B\right) \quad (2)$$

where $m$ is the massive of free electron and $\hbar$ is reduced the Plank's constant. The resulting tunneling possibilities at ML phosphorene-Ag, Au, Al, Cu, Ti, Cr, Ni, and Pd are 59%, 61%, 67%, 85%, 100%, 100%, 100%, and 100%, respectively. The metallization between ML phosphorene and Cr, Ti, Ni, and Pd electrodes make electrons injecting from metal to ML phosphorene freely. Tunneling possibilities of weak bonding is smaller than that of strong bonding.

In above energy band method, the electronic properties of the electrode region and the channel are treated separately, ignoring the coupling between the two regions and thus possible Fermi level spinning. A more reliable to study the SBH of a transistor is to use *ab initio* quantum transport simulations within a two-probe model, in which the coupling between the electrode region and the channel are taken into account. The schematic model the ML phosphorene transistor is presented in Figure 6(a). The electrodes adopt the optimized ML phosphorene-metal interfaces according to our DFT results. The zero-bias transmission spectrums of ML phosphorene FETs with the channel length $L = 5$ nm are plotted in Figure 7. The transmission coefficient $T(E)$ at energy $E$ is obtained from the retarded Green's functions:

$$T(E) = G(E)\Gamma^L(E)G(E)^\dagger \Gamma^R(E) \quad (3)$$

where $G(E)$ is the retarded Green Function, $\Gamma^{L/R}(E) = i\left(\Sigma^r_{L/R} - \Sigma^a_{L/R}\right)$ describes the level broadening due to left electrode/right electrodes expressed in terms of the electrode self-energies $\Sigma^r_{L/R}$, which describe the influence of the electrodes on the scattering region. The transport hole (electron) SBH $\Phi^h_{trans}$ ($\Phi^e_{trans}$) can be regarded as the difference from the Fermi level to the CBM (VBM) of channel ML phosphorene. When Au, Cu, Cr, Al, and Ag are chosen as electrodes, *n*-type ML phosphorene FET is formed with $\Phi_{trans}^{e}$ of 0.30, 0.34, 0.37, 0.51, and 0.52 eV. While when Ti, Ni, and Pd are chosen as electrodes, *p*-type Schottky ML phosphorene FET is formed with $\Phi_{trans}^{h}$ of 0.30, 0.26, and 0.16 eV, respectively. The transmission gap $E_g^{trans} = \Phi_{trans}^{h} + \Phi_{trans}^{e}$ originates from the transport gap of the ML phosphorene channel. The transport gap $E_g^{trans}$ for Ni, Ti, Cu, Al, Ag, Au, Cr, and Pd as electrodes are 0.65, 1.05, 1.09, 1.11, 1.20, 1.20, 1.26, and 1.34 eV, respectively. These transport gaps are close to the band gap of ML phosphorene except for Ni electrode, where short channel effects become especially apparent and broadens the conduction band of ML phosphorene (thus decrease the transport gap) significantly .

We calculated the LDDOS in the ML phosphorene FETs with Au, Al, Ni, and Pd electrodes and show them in Figure 8. ML phosphorene form *n*-type contact with Al and Cu electrode with electron SBHs of 0.52 and 0.31 eV, and *p*-type contact with Ni and Pd electrodes, with hole SBHs of 0.34 and 0.05 eV. The transport gap of ML phosphorene extracted from the LDDOS with Ni, Cu, Al, and Pd electrodes is 0.79, 0.93, 1.12, and 1.13 eV, respectively. The SBHs and transport gaps from the LDDOS calculation are consistent with those from the corresponding transmission spectrum calculations.

Band bending away from the contact is an important property in a metal-semiconductor interface. In generally, the conduction band is bent downward due to an electron transfer from electrodes to channel ML phosphorene for *n*-type contact, while the valence band is bend upward due to an inverse electron transfer in *p*-type contact except for Pd-ML phosphorene interface where the hole Schottky barrier shape looks like a conventional one. Such a band bending picture in a metal-2D semiconductor interface where no impurity state exists in 2D semiconductor (lack of proper doping approach for 2D semiconductor)[23] appears different from a interfacial band bending picture in a common metal-*n* (*p*) type semiconductor interface where donor (acceptor) states exist and electrons (hole) are transferred from semiconductor to metal.

The calculated Schottky barriers in the two methods (the electronic band structure calculations and the quantum transport simulations) are compared in Figure 9. ML phosphorene FET with Al and Ag electrode form the same *p*-type FET in the two methods with similar hole

SBHs (0.68 and 0.61 eV, respectively, from the electronic band structure calculations vs 0.60 and 0.68 eV from the quantum transport simulations). When Ti, Cr, Au, and Cu are used as electrodes, the conduction carriers of ML phosphorene FET are completely reversed in the two methods. ML phosphorene FET with Ti electrode changes from *n*-type Schottky contact with a larger hole SBH of 0.51 eV (the electronic band structure calculations) to *p*-type Schottky contact with a much smaller hole SBH of 0.30 eV (the quantum transport simulations) FET. The available experimental results shows that phosphorene FET with Ti electrode forms *p*-type Schottky FET, and few layer phosphorene with Ti electrode have a hole SBH of 0.21 eV.[13] This result is in favor of the quantum transport simulations.[15] The smaller hole SBH obtained in experiment than obtained in the transport simulation is ascribed to a difference in layer number. Few layer phosphorene has a smaller band gap than ML phosphorene and generally has a smaller hole SBH. By contrast, ML phosphorene FET with Cr, Au, and Cu electrodes change from *p*-type Schottky (the electronic band structure calculations) to *n*-type Schottky (the quantum transport simulations) FET. The hole SBH of ML phosphorene with Ni electrode from the transport simulation (0.26 eV) is apparently larger than that (0.02 eV) from the energy band analysis. The extracted experimental transport hole SBH of ML phosphorene with Ni electrode is 0.35 $\pm$ 0.02 eV, apparently preferring the quantum transport simulation result once more.

ML phosphorene forms Ohmic contact with Pd electrode in the electronic band energy calculation. However, a small hole SBH (0.05 (LDDOS) ~ 0.16 (Transmission spectrum) eV) appears in quantum transport simulations, which reflects a Fermi level pinning and is line with that the observation that phosphorene forms *p*-type FET with the smallest contact resistance value among experimental checked Ti, Ni, and Pd electrodes.[15,46] Artificial Ohmic contact is also calculated for 2D $MoS_2$-Sc contact in the energy band analysis as a result of ignoring the coupling between the electrode and channel region.[27] This deficiency is corrected by the quantum transport simulations. From the above comparison, the quantum transport simulations are more accurate to evaluate the Schottky barriers at ML phosphorene FET than the electronic band energy calculations. What's more, there are obvious Fermi level pinning between the metalized ML phosphorene electrodes and channel ML phosphorene in all checked ML phosphorene FET.

According to the tunneling barrier and transport Schottky barriers in the dual interfaces model, three kinds of ML phosphorene-metal contacts are identified and depicted in Figure 6b-d. ML phosphorene forms *n*-type Schottky contact with Al, Ag, Au, and Cu electrodes in the interface D with a tunneling barrier in the interface B, leading to type 1 contact. In type 2 contacts, ML phosphorene-Cr system forms a *n*-type Schottky contact at the interface D without a tunneling barrier in the interface B. In type 3 contact, ML phosphorene-Ti, Ni, and Pd systems form a *p*-type Schottky contact at the interface D without a tunneling barrier in the interface B. In all the ML phosphorene-metal systems, no SBH is formed in the interface B.

In summary, we have investigated the interfacial bonding level and contact barriers in ML phosphorene and commonly used metal electrodes Al, Ag, Cu, Au, Cr, Ni, Ti, and Pd interfaces by using *ab initio* electronic structure calculations and quantum transport simulations. Unlike graphene[38,39], graphdiyne[24], and TMDs[27] strongly interacted with the partial checked metals, ML phosphorene strongly interact with all the checked metals. Compared with the energy band calculations, the quantum transport simulations are more reliable in describing SBH because of including the interactions between metal electrodes and channel ML phosphorene and better consistent with the experimental results. In terms of the quantum transport simulations, the previously reported Ohmic contact between Cu/Pd and ML phosphorene contacts turns out to be artificial. *N*-type Schottky contacts are formed between ML phosphorene and Au, Cu, Cr, Al, and Ag electrodes, with electron SBHs of 0.30, 0.34, 0.37, 0.51, and 0.52 eV, respectively, and *p*-type Schottky contacts are formed between ML phosphorene and Ti, Ni, and Pd electrodes, with hole SBHs of 0.30, 0.26, and 0.16 eV, respectively. These results are well supported by the experimental data. The further design of ML phosphorene electronic devices can reference our results to choose an appropriate metal electrode.

**Acknowledgement**

This work was supported by the National Natural Science Foundation of China (No. 11274016), the National Basic Research Program of China (No. 2012CB619304/No. 2013CB932604), Fundamental Research Funds for the Central Universities, National Foundation for Fostering Talents of Basic Science (No. J1030310/No. J1103205), Program for New Century Excellent Talents in University of MOE of China.

**Table 1** Calculated interfacial properties of ML phosphorene-metal contacts. $\bar{\varepsilon}$ is the average absolute ML phosphorene surface lattice constant mismatch. The equilibrium distance $d_{P-M}$ is the average distance between the contact ML phosphorene-metal interfaces in the vertical direction of it. The minimum interatomic distance $d_{min}$ is the minimum atomic distance from the innermost layer atom of metal to ML phosphorene surface atom. The binding energy $E_b$ is the energy of per phosphorus atom to remove ML phosphorene from metal surface. $W_M$ and $W$ are the calculated work function for clean metals surface and adsorbed ML phosphorene-metal system, respectively. $\Delta V$, $w_B$, and $T_B$ are the tunneling barrier height, tunneling barrier width, and tunneling possibility, respectively. $\Phi_L^h$ ($\Phi_L^e$) and $\Phi_{trans}^h$ ($\Phi_{trans}^e$) are the electronic and transport SBH of hole (electron) in the lateral direction. $E_g^{trans}$ is the transport gap, define as $E_g^{trans} = \Phi_{trans}^e + \Phi_{trans}^h$. The calculated work function of ML phosphorene is $W_P = 5.04$ eV.

| | Al | Ag | Cu | Au | Ti | Cr | Ni | Pd |
|---|---|---|---|---|---|---|---|---|
| $\bar{\varepsilon}$ (%) | 0.52 | 1.12 | 2.58 | 1.00 | 2.58 | 1.02 | 2.65 | 0.96 |
| $d_{P-M}$ (Å) | 2.46 | 2.44 | 2.30 | 2.38 | 1.78 | 1.69 | 1.77 | 2.19 |
| $d_{min}$ (Å) | 2.55 | 2.57 | 2.35 | 2.46 | 2.32 | 2.23 | 2.21 | 2.28 |
| $E_b$ (eV) | 0.41 | 0.43 | 0.59 | 0.46 | 0.89 | 1.62 | 1.32 | 0.85 |
| $W$ (eV) | 4.36 | 4.43 | 4.89 | 4.84 | 4.53 | 4.71 | 5.02 | 5.32 |
| $W_M$ (eV) | 3.86 | 4.20 | 4.77 | 5.00 | 4.35 | 4.69 | 5.16 | 5.27 |
| $\Delta V$ (eV) | 1.26 | 1.67 | 0.65 | 1.55 | 0.00 | 0.00 | 0.00 | 0.00 |
| $w_B$ (Å) | 0.69 | 0.80 | 0.38 | 0.76 | 0.00 | 0.00 | 0.00 | 0.00 |
| $T_B$ (%) | 0.67 | 0.59 | 0.85 | 0.61 | 100.00 | 100.00 | 100.00 | 100.00 |
| $\Phi_L^e$ (eV) | 0.23 | 0.30 | 0.76 | 0.71 | 0.40 | 0.58 | 0.89 | 0.00 |
| $\Phi_{trans}^e$ (eV) | 0.51 | 0.52 | 0.34 | 0.30 | 0.75 | 0.37 | 0.39 (0.64) | 1.18 |
| $\Phi_L^h$ (eV) | 0.68 | 0.61 | 0.15 | 0.20 | 0.51 | 0.33 | 0.02 | 0.00 |
| $\Phi_{trans}^h$ (eV) | 0.60 | 0.68 | 0.75 | 0.90 | 0.30 | 0.89 | 0.26 (0.35) | 0.16 |
| $E_g^{trans}$ (eV) | 1.11 | 1.20 | 1.09 | 1.20 | 1.05 | 1.26 | 0.65 (0.99) | 1.34 |

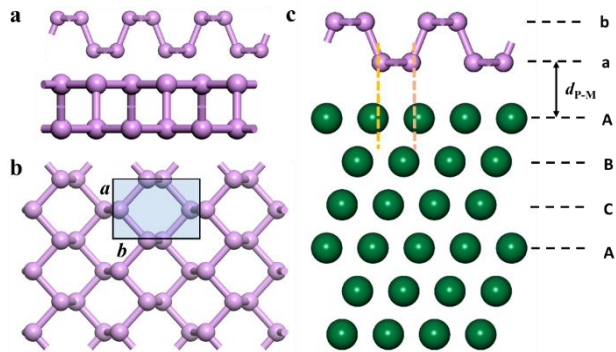

**Figure 1.** (a) Side and (b) top views of freestanding ML phosphorene. The rectangle indicates the unite cell of ML phosphorene. (c) The initial configuration of ML phosphorene (purple ball) on metal surface (green ball).

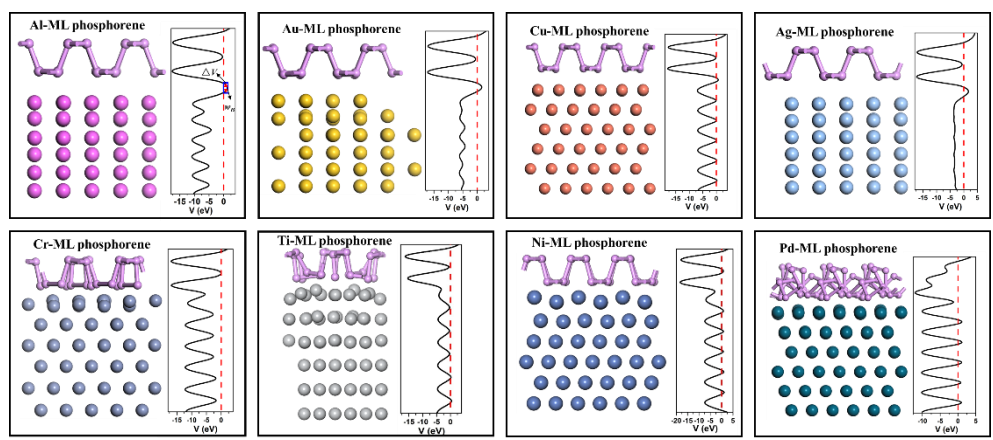

**Figure 2.** Side view of the optimized structures and average electrostatic potentials in planes normal to the interface of ML phosphorene-Al, Au, Ti, Cr, Ag, Cu, Ni, and Pd systems, respectively. The Fermi level is set to zero.

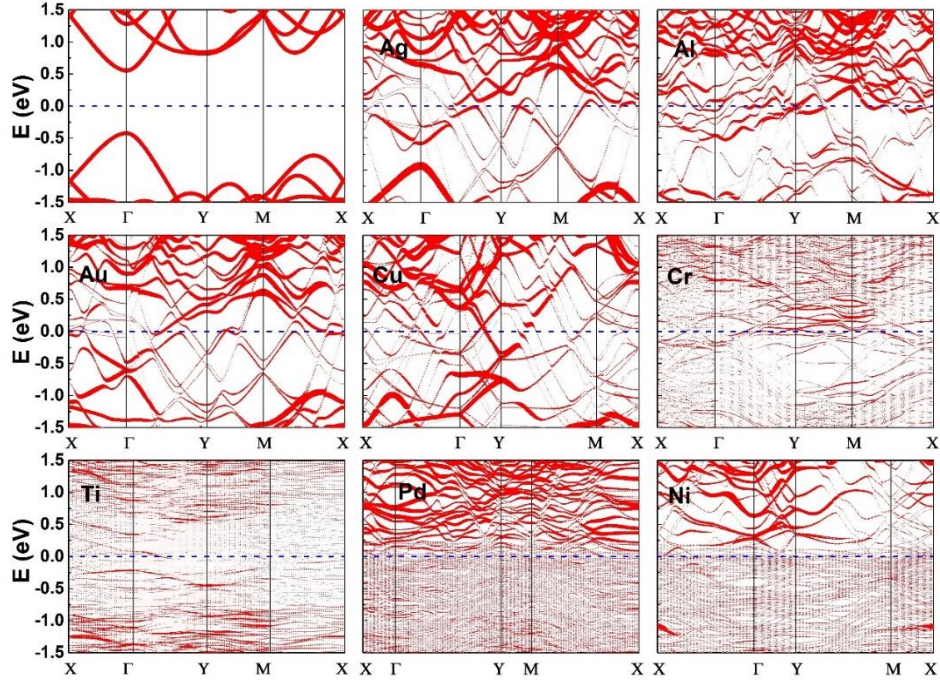

**Figure 3.** Band structures of ML phosphorene and ML phosphorene on Ag, Al, Au, Cu, Cr, Ti, Pt, and Ni surfaces by the DFT method, respectively. The Fermi level is at zero energy. Gray line: band structure of the interfacial systems; red line: band structures of ML phosphorene. The line width is proportional to the weight. The band structure of freestanding ML phosphorene is calculated in $\sqrt{a^2+b^2}\times\sqrt{(2a)^2+b^2}$ supercell.

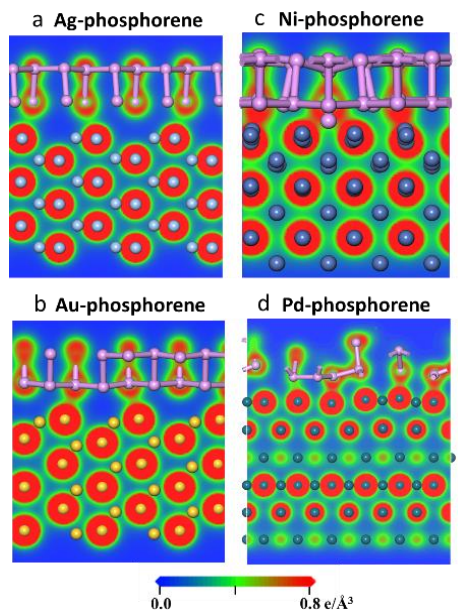

**Figure 4.** Contour plots of total electron distribution of a. ML phosphorene-Ag system, b. ML phosphorene system, c. ML phosphorene-Ni system, and d. ML phosphorene-Pd systems. The purple, light blue, dark blue, yellow and green balls represent P, Ag, Ni, Au, and Pd atoms, respectively.

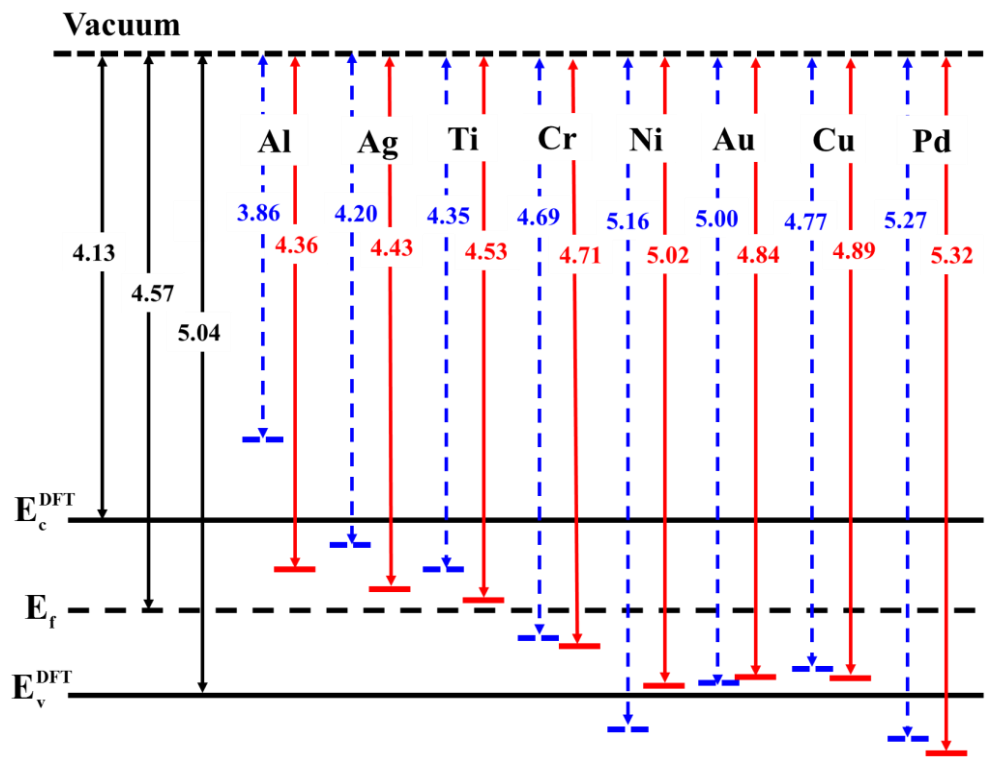

**Figure 5.** Line-up of the work functions of the interfacial systems with the electronic band of channel ML phosphorene in terms of separate the electronic energy band calculations. The blue dash and the red solid present the work function of the pure metals and ML phosphorene-metals systems, respectively.

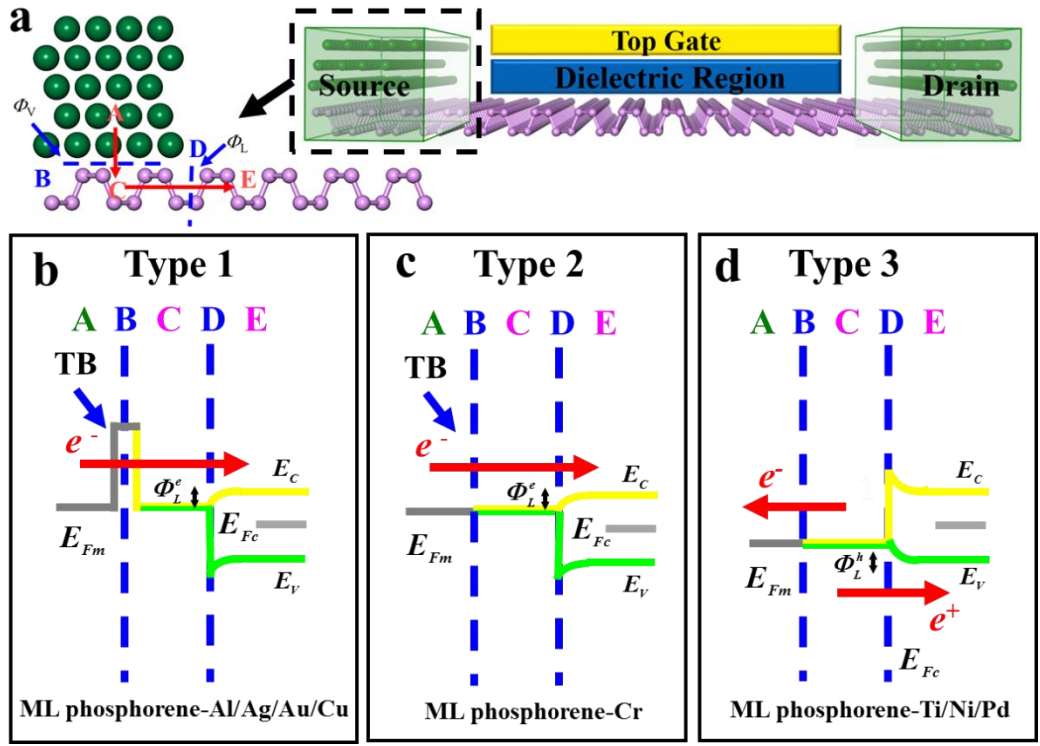

**Figure 6.** (a) Right: Schematic diagram of a ML phosphorene FET. Left: Schematic cross-sectional view of a typical metal contact to intrinsic ML phosphorene channel. A, C, and E denotes three regions, while B and D are the two interfaces separating them. Red rows show the pathway (A→B→C→D→E) of electron or hole injection from contact metal (A) to the ML phosphorene channel (E). (b-d). Three possible band diagrams of the ML phosphorene FET, depending on the type of metal. Examples are provided at the bottom of each diagram. $E_{Fm}$ and $E_{Fc}$ denote the Fermi level of the interfacial system and channel ML phosphorene, respectively.

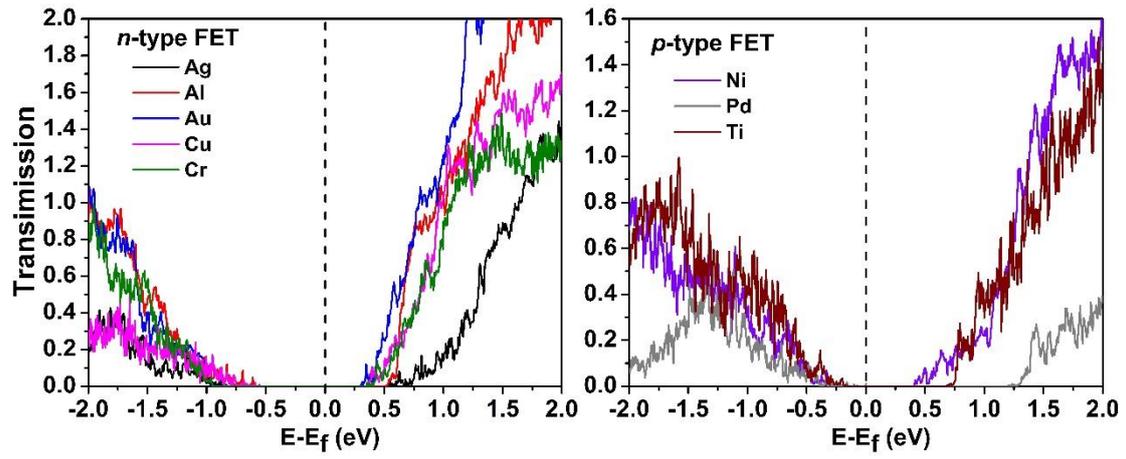

**Figure 7.** Zero-bias transmission spectra of the channel in ML phosphorene FET with Ag, Cu, Cr, Al, Au, Ni, Ti, and Pd electrodes, the channel length $L = 5$ nm.

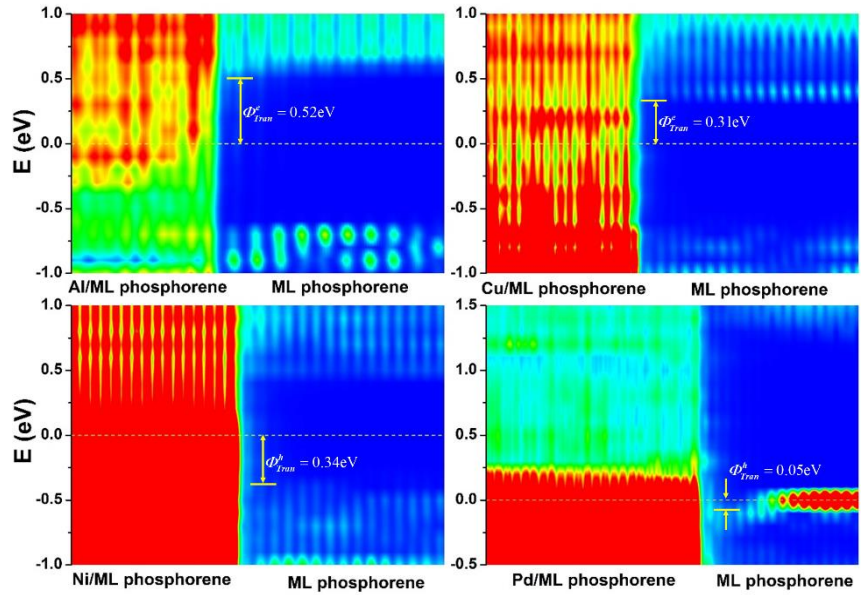

**Figure 8.** LDDOS in color coding for ML phosphorene FET with Al, Cu, Ni, and Pd electrodes, respectively.

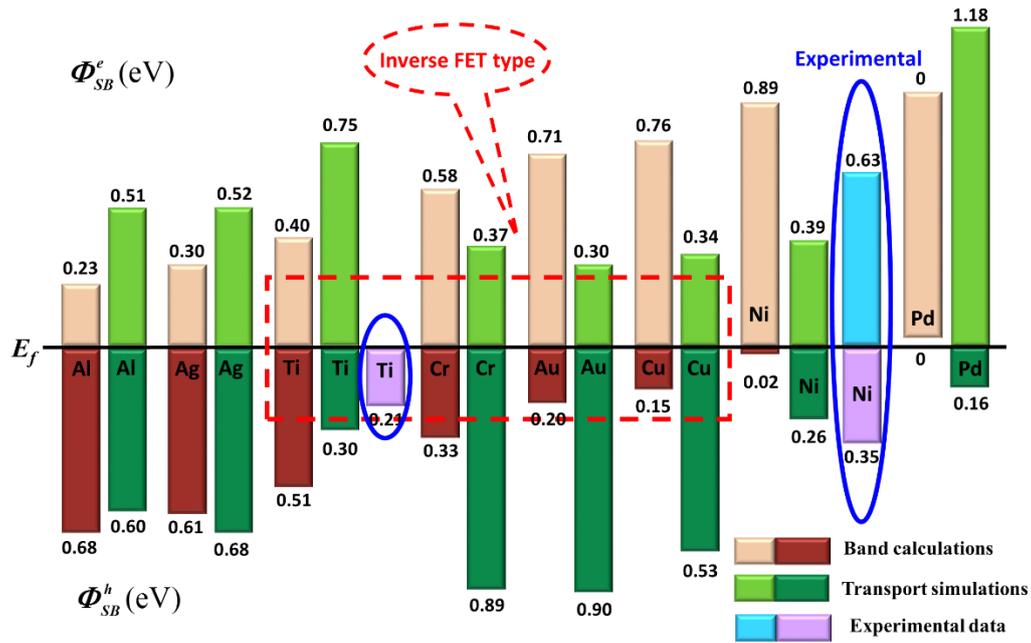

**Figure 9.** Comparison of the Schottky barriers obtained by the *ab initio* electronic band calculations, quantum transport calculations, and experimental observations.[13,14] $\Phi_{SB}^{e}$ and $\Phi_{SB}^{h}$ of ML phosphorene-Al, Ag, Ti, Cr, Cu, Au, Ni, and Pd systems.

TOC

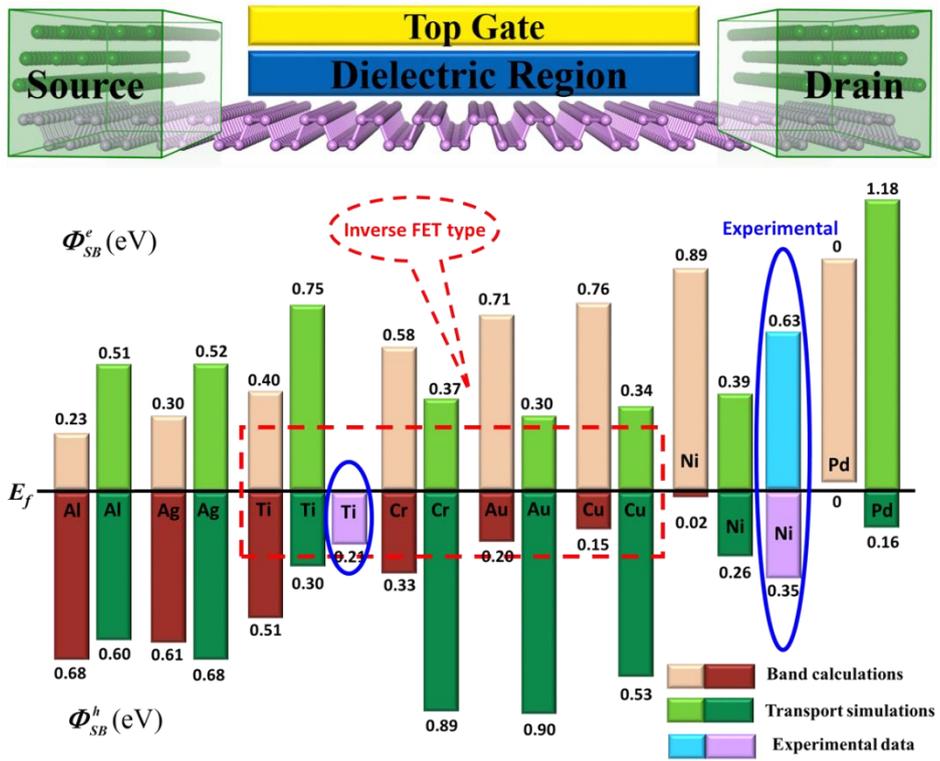